\documentclass[sigconf, natbib=false]{acmart}

\usepackage{subcaption}
\usepackage{listings}
\usepackage{xcolor}
\usepackage{pgfplots}
\usepackage{amsmath, amssymb}

\usepackage{algorithm}
\usepackage{algorithmic}

\pgfplotsset{compat=1.17}
\usepackage[most]{tcolorbox}
\usetikzlibrary{shapes, arrows.meta, positioning}

\usepackage{xcolor} 
\definecolor{myorange}{RGB}{252,194,3}
\definecolor{myblue}{RGB}{4,131,199}
\definecolor{myred}{RGB}{255,103,102}

\definecolor{mycolor1}{RGB}{255, 217, 101}
\definecolor{mycolor2}{RGB}{14,128,136}
\definecolor{mycolor3}{RGB}{151,114,166}

\RequirePackage[
  datamodel=acmdatamodel,
  style=acmnumeric,
  sorting=none
  ]{biblatex}
\begin{document}

\title{PenHeal: 
A Two-Stage LLM Framework for Automated Pentesting and Optimal Remediation}

\author{Junjie Huang}
\email{jh7956@nyu.edu}
\affiliation{%
  \institution{New York University Shanghai}
  \city{Shanghai}
  \country{China}
}

\author{Quanyan Zhu}
\email{quanyan.zhu@nyu.edu}
\affiliation{%
  \institution{New York University}
  \city{New York}
  \country{USA}
}

\renewcommand{\shortauthors}{Huang et al.}

\begin{abstract}
Recent advances in Large Language Models (LLMs) have shown significant potential in enhancing cybersecurity defenses against sophisticated threats. LLM-based penetration testing is an essential step in automating system security evaluations by identifying vulnerabilities. Remediation, the subsequent crucial step, addresses these discovered vulnerabilities. Since details about vulnerabilities, exploitation methods, and software versions offer crucial insights into system weaknesses, integrating penetration testing with vulnerability remediation into a cohesive system has become both intuitive and necessary.

This paper introduces \textit{PenHeal}, a two-stage LLM-based framework designed to autonomously identify and mitigate security vulnerabilities. The framework integrates two LLM-enabled components: the Pentest Module, which detects multiple vulnerabilities within a system, and the Remediation Module, which recommends optimal remediation strategies. The integration is facilitated through Counterfactual Prompting and an Instructor module that guides the LLMs using external knowledge to explore multiple potential attack paths effectively. Our experimental results demonstrate that \textit{PenHeal} not only automates the identification and remediation of vulnerabilities but also significantly improves vulnerability coverage by 31\%, increases the effectiveness of remediation strategies by 32\%, and reduces the associated costs by 46\% compared to baseline models. These outcomes highlight the transformative potential of LLMs in reshaping cybersecurity practices, offering an innovative solution to defend against cyber threats.
\end{abstract}




\maketitle

\section{Introduction}

The rapid evolution of large language models (LLMs) such as OpenAI's GPT-3 \cite{brown2020language} and GPT-4 \cite{openai2024gpt4}, Meta's Llama 2 \cite{touvron2023llama}, and Google's Gemini \cite{geminiteam2024gemini}
showcases their expanding role across a spectrum of natural language processing tasks. 
These models, equipped with sophisticated architectures and extensive learning capacities,
excel in a wide range of tasks including text generation \cite{textgenSurvey}, translation \cite{wang2019learningdeeptransformermodels}, 
summarization \cite{zhang2019pretrainingbasednaturallanguagegeneration}, and so on,
highlighting their potential to innovate across various domains.

Meanwhile, in the digital era, the severity and frequency of cyber attacks are on the rise, 
posing a growing threat to the security of critical systems and data belonging to numerous individuals and organizations \cite{cyberattack}. 
These malicious activities inflict immediate losses
potentially triggering profound socio-economic consequences.
Against this backdrop, penetration testing emerges as a crucial defensive strategy \cite{AboutPenetrationTesting}.
By simulating cyber attacks, penetration testing proactively identifies vulnerabilities within systems before they are exploited by malicious entities, and then the vulnerabilities discovered could be remediated by cybersecurity professionals.

However, the traditional manual operations involved in penetration testing are often time-consuming and heavily reliant on specialized expertise. 
To address these challenges, LLMs are being increasingly integrated into cybersecurity practices, tackling tasks such as vulnerability detection in source code \cite{zhou2024large}, 
threat intelligence \cite{shafee2024evaluation}, and system log analysis \cite{omar2023detecting}. 
PentestGPT \cite{deng2023pentestgpt}, incorporating multiple LLMs, can successfully guide human operators through penetration testing. AUTOATTACKER \cite{xu2024autoattacker} achieves complete automation in post-breach scenarios, demonstrating the effectiveness of LLMs in separated security tasks. 
Nonetheless, these approaches either involve human intervention or are limited to narrowly focused tasks, highlighting the challenge of developing LLMs that can autonomously navigate diverse cybersecurity environments and efficiently identify multiple vulnerabilities concurrently. 

Furthermore, details about a system's vulnerabilities (for example, exploitation methods and software versions) offer critical insights into existing weaknesses and their corresponding remediation strategies, so it is intuitive and essential to integrate penetration testing and vulnerability remediation into a unified system. Yet, despite the advancements in the automated detection of vulnerabilities, 
the area of automated vulnerability remediation is still largely unexplored. 
This represents a critical research gap as organizations increasingly confront cyber threats that demand not only exhaustive identification but also effective and efficient resolution. 

Given these challenges, we aim to investigate the following \textbf{research questions}:
\begin{itemize}
  \item \textbf{RQ1}: Can LLMs automate penetration testing that discovers \textbf{multiple} vulnerabilities in a target system and \textbf{without human intervention}?
  \item \textbf{RQ2}: Can LLMs automate vulnerability remediation \textbf{effectively} and \textbf{cost-efficiently}?
\end{itemize}

Our proposed model, \textbf{\textit{PenHeal}},
includes two main components: 
the \textbf{Pentest Module} and the \textbf{Remediation Module}. 
The Pentest Module is responsible for detecting vulnerabilities, and this process is facilitated by \textbf{Counterfactual Prompting} to explore multiple attack paths and the \textbf{Instructor} module to guide the model with external knowledge.
Meanwhile, the Remediation Module translates vulnerabilities detected by the Pentest Module into actionable remediation strategies, with the help of the \textbf{Adviser} LLM and \textbf{Evaluator} LLM that generate and select optimal remediation suggestions. 
Both components, innovatively integrating multiple LLM components, work together to enhance the coverage rate of penetration testing together with the efficiency and effectiveness of vulnerability remediation with full automation. 

Our \textbf{contributions} in this work are as follows:
\begin{itemize}
  \item We design the Pentest Module capable of autonomously identifying multiple vulnerabilities in a target system with full automation.  
  \item We conduct the first study to assess the capabilities of LLMs in automating vulnerability remediation 
        and design the Remediation Module that generates high-quality remediation recommendations for the detected vulnerabilities.
  \item We establish a new benchmark to evaluate: (1) the penetration testing's coverage rate of vulnerabilities; (2) the effectiveness and efficiency of vulnerability remediation.
\end{itemize}

\section{Background and Related Works}


\subsection{Cybersecurity Frameworks}

\subsubsection{Penetration Testing}

Penetration testing, or ``pentest'', is a critical strategy for enhancing system security.
It involves a simulated attack against a computer system to identify exploitable vulnerabilities. 
During penetration testing, pentesters would try to hack the system as a malicious attacker would, conducting steps that enable them to pinpoint vulnerabilities within the system: Reconnaissance, Scanning, Vulnerability Assessment, Exploitation, and Post-Exploitation \cite{weidman2014penetration}. 

Yet, because of the repetitive and time-consuming nature of penetration testing, 
recent research has begun to explore the potential of transition from manual to automated penetration testing using AI-enabled tools.
Several studies have proposed the use of Reinforcement Learning (RL) to develop penetration testing tools, 
employing techniques such as Q-learning and Deep Q-learning (DQN) \cite{hu2020, chu2018, chaudhary2020}. 
However, the test cases in these studies are often simplistic or artificially constructed, 
and the models tend to underperform in high-dimensional environments with large action spaces. 
This highlights the need for further advancements in AI methodologies to enhance the effectiveness and scalability of automated penetration testing tools.

\subsubsection{Vulnerability Remediation}
Current research on AI-based vulnerability remediation predominantly centers on addressing vulnerabilities in source code \cite{fu2022vulrepair,marchandmelsom2020automatic,pinconschi2021comparative}, involving approaches such as neural transfer learning \cite{Chen_2023} and 
generative adversarial networks \cite{harer2018learning}.
These approaches show success in detecting and mitigating coding flaws effectively. 
However, despite these advancements, there remains a notable gap in the research 
concerning the remediation of vulnerabilities that affect computer hosts, systems, and network infrastructures.

\subsubsection{CVE and CVSS}
In cybersecurity, the Common Vulnerabilities and Exposures (CVE) system \cite{CVE} plays a 
pivotal role by providing a standardized identifier for known security vulnerabilities.
Each CVE entry details a specific vulnerability, 
enabling cybersecurity professionals to communicate about security threats using a unified language. 
Complementing CVE, Common Vulnerability Scoring System (CVSS) \cite{cvss31spec} 
rates the severity of vulnerabilities based on their impact and complexity, helping prioritize and strategize remediation efforts. In our research, we utilize the CVE and CVSS frameworks to enhance the accuracy of the remediation recommendations generated by our model.

\subsection{LLMs and Their Cybersecurity Application}

\subsubsection{Recent development of LLMs}
In recent years, the landscape of natural language processing has been transformed by the rapid advancement of Large Language Models (LLMs). 
Powered by the transformer architecture \cite{vaswani2023attention}, 
Prominent models such as GPT-3 \cite{brown2020language}, GPT-4 \cite{openai2024gpt4}, and Llama 2 \cite{touvron2023llama}, Google Gemini \cite{geminiteam2024gemini} have set new benchmarks in a variety of tasks: text generation \cite{textgenSurvey}, translation \cite{wang2019learningdeeptransformermodels}, 
summarization \cite{zhang2019pretrainingbasednaturallanguagegeneration}, question-answering \cite{namazifar2021}, etc. 
In this paper, we focus on utilizing OpenAI's GPT models, specifically GPT-3.5 and GPT-4, as the foundation model to explore the application of LLMs in automated penetration testing and vulnerability remediation. 

\subsubsection{LLM enhancing techniques}

LLMs can store vast amounts of factual knowledge,
yet they often struggle with generating responses that are consistently accurate and up-to-date, 
especially in rapidly evolving fields or niche areas where specific expertise is required.
To address this, Retrieval-Augmented Generation (RAG) models \cite{lewis2021retrievalaugmented} 
enhance LLM capabilities by merging pre-trained parametric memory with dynamic non-parametric memory systems stored as densely packed semantic vectors. 
When a user inputs a query, the RAG model activates its retrieval mechanism to find and utilize the most relevant information 
from its knowledge base by comparing vector similarities, enabling the creation of informed responses. 
In specialized domains like penetration testing and remediation, 
this technique can significantly enhance the model's applicability. RAG model is also deployed in our model to enhance its expertise in penetration testing.

Recent research has also shown that role-play prompting 
enhances LLMs' performance in complex tasks by aligning their responses with assigned roles \cite{kong2024better}. 
Building on this, our work applies role-play prompting to penetration testing and vulnerability remediation, 
encouraging LLMs to generate accurate and contextually appropriate responses.

\subsubsection{LLM-based agents}
The strong autonomy of LLMs has established them as popular choices for developing intelligent agents 
capable of interacting with the world using various tools \cite{xi2023rise}. 
Research has shown that LLMs can adeptly learn to use external tools, such as APIs, to efficiently complete tasks \cite{qin2023toolllm} and that multiple LLM-based agents can cooperate to tackle complex tasks \cite{xi2023rise}. 
Our model, \textit{PenHeal}, incorporates several LLM-based agents, 
enabling it to utilize common cybersecurity tools like Metasploit, Nmap, and API access 
to complete cybersecurity tasks.

\subsubsection{LLM's application in cybersecurity}

The application of LLMs in cybersecurity expands various domains, 
including vulnerability detection in source code \cite{zhou2024large},
threat intelligence \cite{shafee2024evaluation}, web content filtration \cite{vörös2023web}, 
vulnerability detection in system logs \cite{omar2023detecting}, 
and conversational security support \cite{kaheh2023cybersentinelexploringconversational}, etc.

In the specific context of attack automation, recent works have shown that 
LLMs are able to handle simple tasks in Capture the Flag (CTF) games \cite{yang2023language, tann2023using, shao2024empirical}.
Moreover,  PentestGPT \cite{deng2023pentestgpt} is an interactive chatbot consisting of reasoning, 
command generation, and output parsing module, 
which could guide pentesters through various stages of an attack. 
It shows the potential of multi-agent LLMs in automating penetration testing tasks. Further, building on their previous work \cite{Happe2023}, \cite{happe2024llms} proposed a fully automated system 
with enhanced guidance through ``Next-Cmd'', ``Analyse-Result'', and ``Update-State'' prompts. 
Their work shows promising results in automating Linux privilege escalation tasks,
indicating the potential of LLMs in automating cybersecurity tasks.
The AUTOATTACKER model \cite{xu2024autoattacker}, meanwhile,  
utilizes a Summarizer, Planner, and Navigator module alongside a RAG-based experience manager 
that allows the system to learn from past attacks and apply this knowledge to future operations. 
It is tested on isolated post-breach tasks such as ``File Writing'' and ``MySQL Scan'',
showing the potential of LLMs in automating post-breach scenarios.

However, these works either focus on specific tasks or require human intervention, 
and no work has been done to explore a fully automated penetration testing system that can autonomously identify multiple vulnerabilities in a target system.
Furthermore, to the best of our knowledge, there is no work that explores the application of LLMs in fully automated penetration testing in a real-world setting, where
we hope to explore vulnerabilities comprehensively and conduct attacks without human intervention. 
In addition, there is a plethora of work on the remediation or patching of discovered vulnerabilities. Therefore, our work aims to fill this significant research gap.

It is also noticed that the benchmark of \cite{deng2023pentestgpt} is employed on commercial platforms, Vulnhub \cite{vulnhub} and HackTheBox \cite{hackthebox},
which are closed-source, charge-required and therefore may not be suitable for academic research. The benchmark of  \cite{xu2024autoattacker}
focuses on individual tasks, which are not suitable for real-world scenarios with multiple vulnerabilities. Furthermore, to the best of our knowledge, 
there is currently no benchmark that can evaluate the effectiveness of the model's remediation recommendations. Consequently, our work also designs a novel benchmark measuring (1) the vulnerability coverage rate of penetration testing; (2) the effectiveness and efficiency of vulnerability remediation.

\section{Problem Setting}
\label{sec:problem_setting}

In practical cybersecurity scenarios, the challenge extends beyond merely detecting individual vulnerabilities. 
Instead, it encompasses the systematic and efficient discovery of all potential security flaws within a system. 
Additionally, the subsequent remediation of these vulnerabilities is equally crucial, where the goal is to propose effective and cost-efficient strategies to mitigate the identified weaknesses.
Therefore, our research problem is centered on the coverage of vulnerability identification and the effectiveness and efficiency of remediation, as indicated by our research questions.

Therefore, we hope our model to excel in these three dimensions:
\begin{itemize}
  \item \textbf{Detection Coverage}: An ideal model should be able to identify a high percentage of the system's vulnerabilities. 
        This is vital to ensure that no critical weaknesses are overlooked, which could otherwise be exploited by attackers.
  \item \textbf{Remediation Effectiveness}: The effectiveness of the proposed fixes is measured by their ability to prevent exploitation post-remediation, 
        which is the drop in the system's vulnerability score after applying the remediation strategies. 
        This attribute of each remediation strategy is quantified as a ``value'' score, which will be explained in detail in Section~\ref{sec:cost_value}.
  \item \textbf{Remediation Efficiency}: Different from detection efficiency, 
        this aspect focuses on the resource expenditure involved in implementing the remediation recommendations. 
        It considers the time, money, and other resources, evaluating how efficiently the remediation strategies can be executed within operational constraints.
        This attribute of each remediation strategy is quantified as a ``cost'' score, and it will be illustrated in detail in Section~\ref{sec:cost_value} as well. 
\end{itemize}

Therefore, we could formalize our objectives as to:
\begin{enumerate}
  \item \textbf{Maximize the coverage rate of vulnerabilities}
  \[
  \max_{\mathcal{V}} \frac{|\mathcal{V}|}{N}
  \]
  where $\mathcal{V} = \mathcal{PEN}(\text{system})$ denotes the set of vulnerabilities identified in the system by the Pentest Module ($\mathcal{PEN}$), and $N$ represents the total number of vulnerabilities in the system.

  \item \textbf{Maximize the value of selected remediation strategies under a budget constraint of cost:}
  \[
  \max_{\mathcal{R}} \sum_{r \in \mathcal{R}} \text{Value}(r) \quad \text{s.t.} \quad \sum_{r \in \mathcal{R}} \text{Cost}(r) \leq \text{Budget}
  \]
  where:
  \begin{itemize}
      \item $\mathcal{R} = \mathcal{REM}(\mathcal{V}) = \mathcal{REM}(\mathcal{PEN}(\text{system}))$ represents the set of remediation strategies recommended by the Remediation Module ($\mathcal{REM}$).
      \item $\text{Value}(r)$ quantifies the effectiveness of a single recommendation $r$.
      \item $\text{Cost}(r)$ quantifies the cost of a single recommendation $r$.
  \end{itemize}
\end{enumerate}

\section{Methodology} 

\lstset{
  basicstyle=\ttfamily\small, 
  breaklines=false,         
  frame=single,                
  backgroundcolor=\color{gray!10},
  rulecolor=\color{black},       
  framesep=2pt,                  
  frameround=ffff,                
  captionpos=b,                    
}

\subsection{Overview}
\begin{figure}[ht]
  \centering
  \includegraphics[width=\columnwidth]{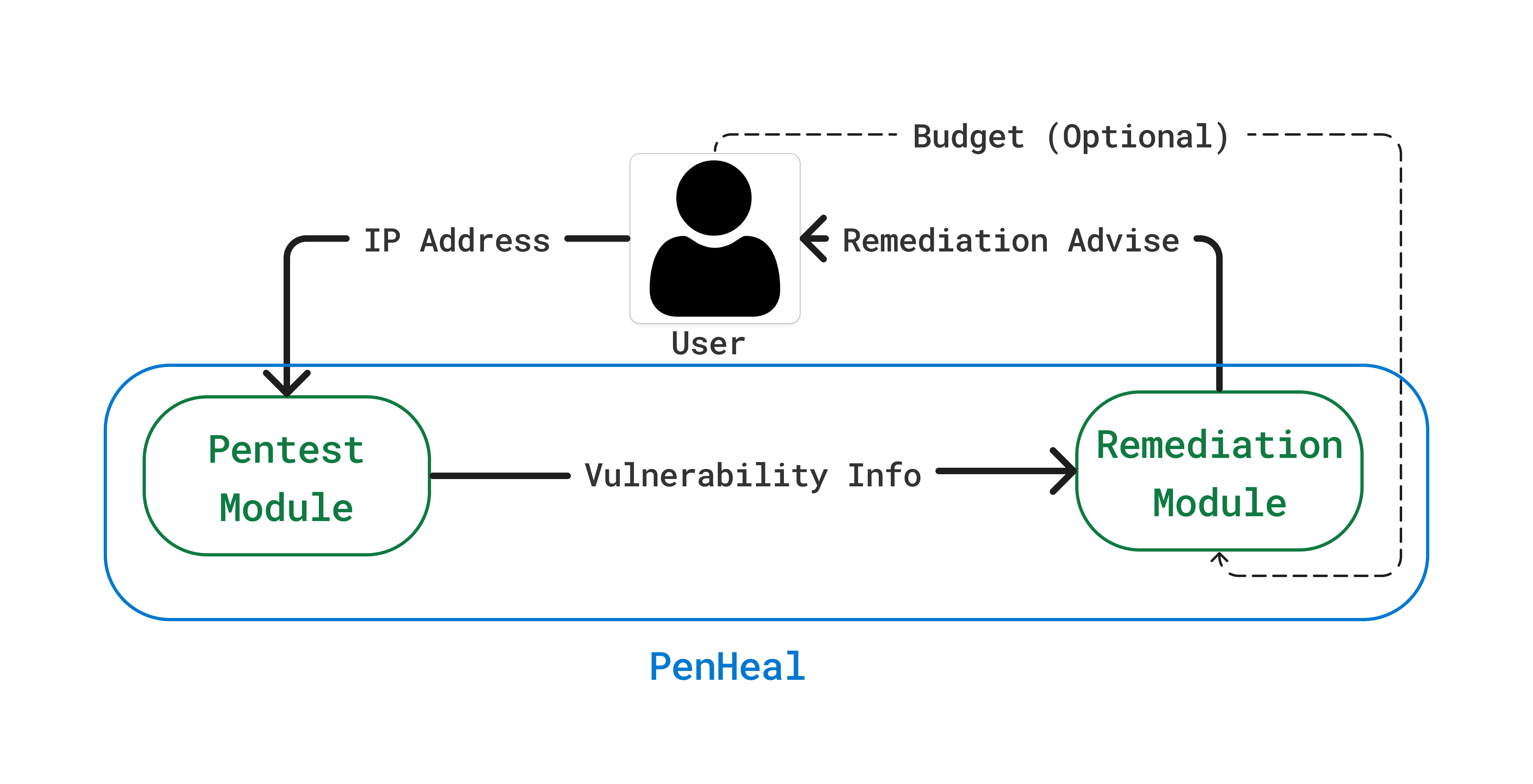}
  \caption{\textit{PenHeal} Abstract Architecture:
     the Pentest Module is responsible for detecting vulnerabilities, and the Remediation Module suggests remediation strategies. Initially, the user inputs the target system's IP address, prompting the Pentest Module to conduct penetration testing and log vulnerabilities. Subsequently, the Remediation Module generates and presents remediation recommendations to the user based on these findings.
  }
  \label{fig:overview}
\end{figure}

Our proposed model, \textit{PenHeal}, integrates two principal components that operate in distinct phases: 
the Pentest Module, tasked with penetration testing discovering multiple vulnerabilities with full automation, 
and the Remediation Module, which is dedicated to generating effective and cost-efficient remediation recommendations. 
The pipeline starts with the user inputting information about the target victim machine (typically the IP address) to the Pentest Module. After that, the Pentest Module scans the system and feeds the discovered vulnerabilities to the Remediation Module. Then, the user can choose to set the cost of different types of remediation approaches and the overall budget in the form of plain text. If not, the default setting is applied and an optimal list of remediation strategies will be presented to the user.
An overview of the model's architecture is shown in Figure~\ref{fig:overview}.

In the following subsections, we will delve into the details of each component and its design rationale.

\subsection{Pentest Module}

\begin{figure*}[ht]
  \centering
  \includegraphics[width=\textwidth]{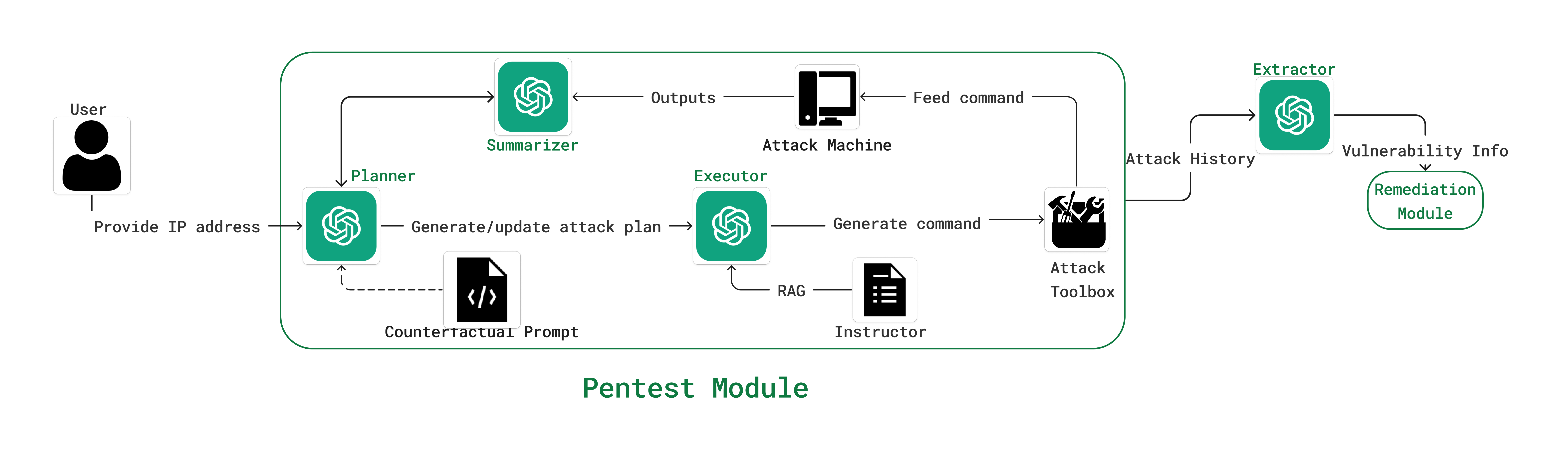}
  \caption{The Pentest Module begins with the user providing the target system's IP address. The Planner then formulates the attack strategy, directing the Executor, which consults the \textbf{Instructor} to generate commands aimed at exploiting system vulnerabilities. These commands are analyzed by the Summarizer, and Counterfactual Prompting aids the Planner in refining the attack strategy by considering alternative vectors in response to new vulnerabilities. The blocks with OpenAI icons represent that they are OpenAI LLMs.}
  \label{fig:pentest_module}
\end{figure*}

The structure of Pentest Module $\mathcal{PEN}$ is inspired by earlier works, PentestGPT \cite{deng2023pentestgpt} 
and AUTOATTACKER \cite{xu2024autoattacker}. The cycle starts with a user inputting the target system's IP address, the only step requiring human intervention. 
Following this, the Planner develops the initial attack strategy and selects a task for the Executor to carry out. Then, the Executor
generates specific commands after consulting the \textbf{Instructor}, which in turn accesses an external knowledge base. 
The command outputs, which may reveal vulnerabilities or results of exploits, 
are then parsed by the Summarizer.  
If new vulnerabilities are detected, Counterfactual Prompting is utilized to encourage the Planner to consider alternative attack vectors, 
thereby refining the ongoing attack strategy based on the latest insights into the victim's environment.
A detailed illustration of the Pentest Module's workflow is shown in Figure~\ref{fig:pentest_module}.

\subsubsection{Planner and Attack Plan}

The Planner LLM, central to our model's penetration testing workflow, functions as the central processor, orchestrating the entire attack process and assigning specific tasks to the Executor to carry out. 
This foundational data structure used by the Planner, the attack plan, is inspired by the Pentesting Task Tree (PTT) \cite{deng2023pentestgpt} and arranges tasks into a structured sequence of reconnaissance, scanning, vulnerability assessment, and exploitation, as outlined in standard procedures \cite{weidman2014penetration}.
The Planner dynamically updates the attack strategy by categorizing tasks into states of completed, to-do, or failed, based on their outcomes informed by the Summarizer. Moreover, compared to PTT, our refined approach ensures that after each task update, the Planner not only records but also provides a concise description of the outputs.
This enhancement helps the Planner gain a clearer high-level understanding of the task's context and specifics, which is essential for generating more specific tasks for the Executor. 
Furthermore, this strategy prevents the Executor from duplicating efforts or repeating errors. For instance, the Executor can accurately set the ``\texttt{LHOST}" parameter in Metasploit commands using the previously recorded IP address of the attack machine, thus avoiding repetitive searches for this information.
A visualization of the attack plan is shown in Figure~\ref{fig:attack_plan_tree} in Section~~\ref{sec:attack_plan_demo}, and its original script is shown in Figure~\ref{fig:attack_plan_sequence} in Appendix Section~\ref{sec:attack_plan_script}. An excerpt of the initial prompt fed to the Planner can also be found in Appendix Section~\ref{sec:planner_system_prompt}.

\subsubsection{Counterfactual Prompting}
We have observed that the model is inclined to repetitively exploit the same vulnerabilities. To address this issue, when the summarized results indicate a successful attack on the previous vulnerability, the Planner is prompted counterfactually and told to make the attack plan with the assumption that certain identified vulnerabilities did not exist. In response to this, the Planner is tasked with not only updating the existing attack plan 
but also pushing itself to consider alternative vulnerabilities that remain to be exploited.
This technique is designed to enhance the model's ability to systematically explore multiple vulnerabilities. 
An example of a counterfactual prompt is shown in Listing~\ref{lst:counterfactual_prompt}.

\begin{lstlisting}[caption=Counterfactual Prompt Example,
label=lst:counterfactual_prompt]
  Here is the list of vulnerabilities already 
  identified, please mark them as completed on the 
  attack list. Now exploit the system as if they 
  do not exist: 
  Port 21/ftp: 
    vsFTPd version 2.3.4 backdoor (CVE-2011-2523)
  Port 6667/irc: 
    Trojaned version of UnrealIRCd (CVE-2010-2075)
  Port 80/html: 
    SQL injection points (Unknown CVE)
\end{lstlisting}

\subsubsection{Executor}
The Executor LLM is
tasked with generating commands based on the assignments from the Planner. 
It ensures that commands are formatted to start and end with ``\$'' for easy parsing, 
addressing the challenge of automation when processing verbose and complex responses from the LLM. 
For instance, when given the task ``\texttt{Scan vulnerabilities}'', the Executor will generate the command ``\texttt{\$nmap --script vuln [target IP]\$}''.
Then, it will be executed in the terminal by spawning a subprocess to interact with attack tools like Metasploit.
Besides, the Executor is prompted to avoid overly
lengthy or complicated commands, because these can be error-prone and 
the insights from previous commands often inform the choice of subsequent commands.
An excerpt of the initial system prompt fed to the Executor is shown in Appendix
Section~\ref{sec:executor_system_prompt}.

\subsubsection{Instructor}
Although LLMs demonstrate strong autonomy and reactivity, 
their ability to generate precise commands can be hampered 
by a lack of detailed knowledge about specific attacking tools. 
To mitigate this, the Instructor component is implemented to provide targeted guidance to the Executor leveraging an external knowledge base.
When the Executor receives a task, the condensed vector form of the task description is automatically
fed to the external database. This vector will be compared with the pre-generated vectors of the tasks in the database, 
and the top $k$ vectors that are most similar to the task will be retrieved. 
For instance, when the Executor receives the task 
``\texttt{Exploit Samba smbd 3.X - 4.X on ports 139 and 445}'', the excerpts showing how to exploit it by using ``\texttt{exploit/multi/samba/usermap\_script}''
 will be retrieved.
Finally, these retrieved excerpts will be fed to the Executor for its reference. The prompt template fed from the Instructor to the Executor is shown in Listing~\ref{lst:Instructor_prompt_template}. A sample excerpt retrieved is shown in Listing~\ref{lst:retrieved_excerpt} in Section~\ref{sec:instructor_demo}.

\begin{lstlisting}[caption=Instructor Prompt Template, label=lst:Instructor_prompt_template]
    Here is a brief introduction to the task: 
    {Task Description}. Here is some info 
    from the knowledge base for your reference:
    1. {Excerpt 1}
    2. {Excerpt 2}
    ......
\end{lstlisting}

\subsubsection{Summarizer}
The Summarizer is responsible for summarizing the outputs of the commands executed by the Executor. 
It is designed to condense the verbose outputs into concise and informative summaries, which
addresses the problem of limited context window in LLMs effectively \cite{deng2023pentestgpt, xu2024autoattacker}.

\subsubsection{Extractor}
The Extractor is a rather simple utility LLM that extracts the vulnerabilities and their characteristics from the recorded attack history. For parsing convenience, it is prompted to output the vulnerability information in a fixed format: ``Exploited: [backdoor/CVE ID]''. For vulnerabilities that are not disclosed, ``CVE-NA'' is filled in. Other attributes about each vulnerability are provided immediately following each "Exploited" block.
This extracted information will be fed to the Remediation Module afterward.

\subsection{Remediation Module}

\begin{figure*}[ht]
  \centering
  \includegraphics[width=\textwidth]{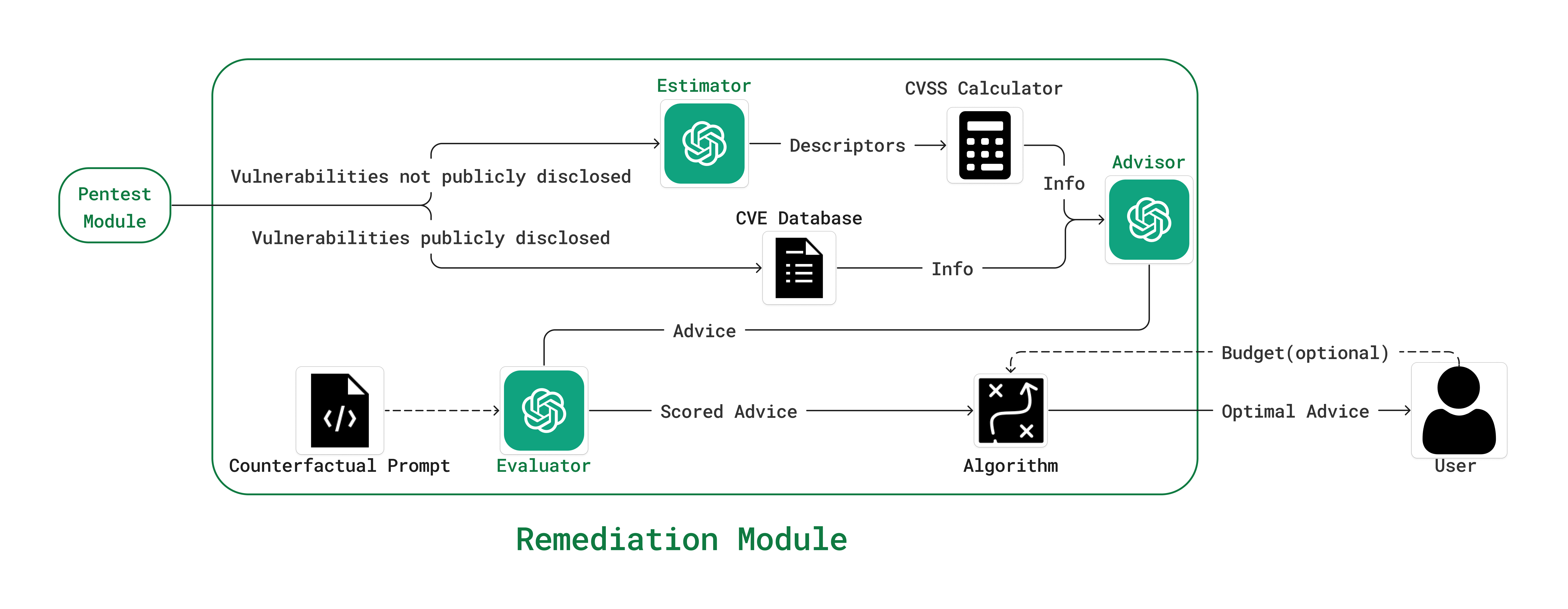}
  \caption{Remediation Module Architecture:
    Upon receiving vulnerability information from the Pentest Module, 
    additional attributes such as CVSS score are retrieved for the vulnerabilities publicly documented. For the others, the Estimator generates descriptors to
    assess their severity and attributes based on the available data. 
    Then, based on the retrieved attributes, the Advisor proposes remediation strategies, 
    which the Evaluator will determine their qualities. Finally, an optimal list of recommendations under the user-defined budget constraint is presented to the user.
  }
  \label{fig:remediation_module}
\end{figure*}

The Remediation Module focuses on formulating remediation strategies for the detected vulnerabilities. 
Among the vulnerabilities received from the Extractor,
for the ones that are already documented (e.g., those with CVE IDs), 
additional information such as CVSS score is retrieved from the database. 
Otherwise, the Estimator will generate descriptors to
assess their severity and attributes based on the available data. 
Then, the Advisor synthesizes this information to propose remediation strategies for the Evaluator to determine their ``cost'' and ``value''. Finally, an optimal list of recommendations that is under the user-tunable budget constraint is selected.
A detailed illustration of the Remediation Module's workflow is shown in Figure~\ref{fig:remediation_module}.

\subsubsection{Estimator and CVSS Lookup}
For vulnerabilities already identified in earlier phases, such as those whose CVE ID is revealed by ``\texttt{nmap --script vuln [target IP]}'', 
our system consults the National Vulnerability Database (NVD) \cite{NVDVulnerabilities} via API requests to gather detailed information. This information includes CVE ID, CVSS score, the description of the vulnerability, the method or backdoor used for exploitation, and etc.
Conversely, for vulnerabilities not yet publicly documented, 
our Estimator LLM will deduce CVSS descriptors based on the available data recorded by the Pentest Module. 
These descriptors are then analyzed by CVSS calculator \cite{cvssPythonPackage} to get their severity scores.
The information collected above is instrumental in determining which vulnerabilities should be prioritized for remediation
and how to best address them, aiding the Advisor module in making informed recommendations. 
An excerpt of the Estimator's initial system prompt is shown in Section~\ref{sec:estimator_system_prompt} in the Appendix.

\subsubsection{Advisor}
The main function of the Advisor LLM is to offer precise, actionable recommendations based on analyzed vulnerabilities. It uses the previously searched attributes to suggest remediation strategies, such as system updates, patches, configuration adjustments, or protective measures. For example, if the Advisor identifies a high-severity vulnerability caused by a backdoor in a particular software version, it may recommend an immediate software update.
Additionally, to boost the effectiveness of the recommendations, the Advisor is prompted to include specific commands or procedures that directly address the identified risks. 
An excerpt of the Advisor's initial system prompt is shown in Appendix Section~\ref{sec:advisor_system_prompt}.

\subsubsection{Evaluator}
\label{sec:cost_value}
The Evaluator LLM is crucial in assessing the remediation recommendations from the Advisor, which, despite being augmented by CVE information, can be numerous and not always pertinent or practical. To aid user decision-making and enhance prioritization, the Evaluator has been integrated with the Group Knapsack Algorithm. This integration allows it to rank recommendations based on a user-tunable quality metric, streamlining the selection process.

As indicated by our RQ2, our goal is to generate remediation recommendations that are both effective and cost-efficient. Therefore, we give every remediation recommendation a ``value'' score based on the CVSS score of the vulnerability it addresses to measure its effectiveness. The rationale behind this is that the higher the CVSS score of a vulnerability, the more severe the vulnerability is,
and the more important it is to remediate it. Similar to the Planner, the Evaluator is prompted counterfactually as well when assigning the recommendations ``value'' scores. It is guided to imagine as if the recommendation is implemented, how effective it could be, and to what degree it could make the system less vulnerable. 
 The Evaluator will assign every recommendation a score based on the following guidelines:

 \begin{itemize}
  \item \textbf{Full CVSS Score:} The full numerical value of the CVSS score is assigned to recommendations that can directly and completely address a vulnerability. For example, a recommendation that successfully patches a vulnerability with a CVSS score of 9.0 will be assigned a ``value'' score of 9.0.
  
  \item \textbf{Partial CVSS Score ($k\%$):} A fraction of the CVSS score, specifically $k\%$, is assigned to recommendations that can only partially address the vulnerability. Here, $k$ is a parameter between 0 and 100, reflecting the effectiveness of the recommendation, with the exact value determined by the Evaluator LLM based on the recommendations.
  
  \item \textbf{Zero Score:} A score of 0 is assigned to recommendations that cannot address the vulnerability or are irrelevant. For instance, a recommendation to change the password of a user account when the vulnerability is due to a service misconfiguration receives a score of 0.
  
  \item \textbf{Sum of Scores for Multiple Vulnerabilities:} If a recommendation addresses multiple vulnerabilities, the ``value'' score assigned is the sum of the CVSS scores of all addressed vulnerabilities.
  
  \item \textbf{Negative Impact (Negative $k\%$):} Negative $k\%$ of the CVSS score is assigned to recommendations that exacerbate the vulnerability. The parameter $k$, ranging from 0 to 100, quantifies the degree to which the recommendation increases the system's vulnerability.
\end{itemize}

As for the ``cost'' scores, users can assign them to remediation recommendations based on their preferences, considering factors like time, money, risk, and other resources. A higher ``cost'' score means the recommendation requires more resources and therefore is less efficient. For example, if users value time highly, they might give a higher ``cost'' score to time-intensive recommendations such as manually writing patch scripts or downloading large software packages. To make the ``cost'' comparable to the ``value'' score, it is defined to fall in the range from 0 to 10 like the ``value'' score.
While this metric is tunable, we provide a default setting for the ``cost'' scores. It also serves as a template for the user to instruct the model about their preferences:

\begin{itemize}
  \item \textbf{Low-Cost recommendations}: These include applying free patches, making configuration changes, or executing simple commands. They typically require minimal resources and are assigned a ``cost'' score of 2.
  
  \item \textbf{Moderate-Cost recommendations}: This category includes recommendations that require manually writing scripts or programs, purchasing software or hardware, or involving some risk. They are assigned a moderate ``cost'' score of 5.
  
  \item \textbf{High-Cost recommendations}: recommendations that necessitate stopping a service, shutting down a system, or carrying a high risk of causing system disruptions fall into this category. Given the significant impact and disruption they may cause, these are assigned the highest ``cost'' score of 10.
\end{itemize}

To simulate the real-world scenario, 
the users are prompted to set a budget for remediation strategies, representing the resources they are willing to allocate. Hence, the objective is to select a set of remediation recommendations that maximize the total ``value'' without exceeding this budget. Additionally, to avoid redundancy, only one recommendation per vulnerability is allowed, ensuring no overlap among chosen strategies. Thus, this can be framed as a Group Knapsack Problem, where each recommendation has associated ``cost'' and ``value'' scores. Employing this algorithm helps identify the optimal solution to our objective.
More explanations of the algorithm are provided in Section~\ref{sec:group_knapsack} in the Appendix. A sample prompt fed to the Evaluator is shown in Appendix Section~\ref{sec:evaluator_system_prompt}.

\section{Experiments}

\subsection{Experimental Setup}
\subsubsection{Environment}
We use the Metasploitable2 Linux virtual machine (2019-08-19 version) as the victim environment. 
Table~\ref{tab:metasploitable_vulnerabilities} shows the summary of all vulnerabilities exploitable from the terminal in Metasploitable2 according to its official documentation \cite{metasploitable2}.
It was chosen because it is an open-source project that is freely available for download, making it accessible to researchers and practitioners.

\begin{table}[h]
  \centering
  \caption{Summary of Vulnerabilities in Metasploitable 2:
  There are 10 types that can be exploited from the terminal.}
  \label{tab:metasploitable_vulnerabilities}
  \begin{tabular}{|c|c|c|}
  \hline
  \textbf{Service} & \textbf{Port} & \textbf{Description} \\ \hline
  FTP (vsFTPd) & 21 & Backdoor shell \\ \hline
  SSH, PostgreSQL, MySQL & 22, 3306, 5432 & Weak passwords \\ \hline
  Telnet & 23 & Default credentials \\ \hline
  SMTP & 25 & Open port \\ \hline
  DNS & 53 & DNS poisoning \\ \hline
  HTTP & 80 & Web vulnerabilities \\ \hline
  NFS & 2049 & Root access \\ \hline
  IRC (UnrealIRCd) & 6667 & Backdoor command \\ \hline
  Samba & Various & Symlink traversal \\ \hline
  PHP CGI scripts & 80 & Argument injection \\ \hline
  \end{tabular}
\end{table}

The attacking host machine in our setup is a Kali Linux virtual machine (2024-01 version), 
equipped with essential penetration testing tools such as the Metasploit Framework and Nmap. It is connected to the same private network as the victim machine, allowing direct communication.

\subsubsection{Use of LLM} 
\label{sec: LLM_user}
It has been established from previous research \cite{deng2023pentestgpt, xu2024autoattacker} that GPT-4 outperforms GPT-3.5 and other publicly available LLMs in penetration testing tasks. 
Consequently, we utilize GPT-4 with a context window size of 128,000 tokens 
for components requiring robust reasoning capabilities, such as the Planner, Executor, Advisor, and Evaluator. 
For components with lesser reasoning demands and functioning primarily as utility modules, namely the Summarizer, Extractor, and Estimator, 
we employ GPT-3.5-turbo with a reduced context window size of 16,000 tokens to minimize computational costs. 
This strategic deployment of LLMs ensures optimal performance while reducing computational costs. 

For the Instructor component, our RAG model, 
the popular Langchain \cite{langchain2023} architecture is used to support the interaction between the LLMs and the external knowledge base.
We have selected two important sources as our external knowledge base: 
\begin{itemize}
  \item \textit{Penetration Testing: A Hands-On Introduction to Hacking} \cite{weidman2014penetration}:
  This book provides a comprehensive introduction to penetration testing, encompassing the full spectrum of the field.
  \item \textit{Metasploit Penetration Testing Cookbook} \cite{jaswal2018metasploit}:
  As a practical guide, this book details the use of Metasploit, 
  a crucial tool in the penetration testing arsenal, and it provides step-by-step instructions across various scenarios.
\end{itemize}

\subsection{Benchmark}

We design a scoring system to evaluate the performance of our model from three aspects mentioned in our Section~\ref{sec:problem_setting}:
\begin{itemize}
  \item \textbf{Detection Coverage Score ($S_{D}$)}: This score is 10 times the percentage of vulnerabilities detected by the model out of the total number of vulnerabilities in the Metasploitable2 environment. The factor of 10 normalizes this score to be on a scale from 0 to 10, making it comparable with other scores.
        \[
        S_{D} = 10 \times \frac{|\mathcal{V}|}{N}
        \]
          where $\mathcal{V} = \mathcal{PEN}(\text{system})$ denotes the set of vulnerabilities identified in the system by the Pentest Module ($\mathcal{PEN}$), and $N$ represents the total number of vulnerabilities in the system.
  \item \textbf{Remediation Effectiveness Score ($S_{R}$)}: Measures the effectiveness of the remediation strategies, or the drop in the system's total vulnerability score after applying them. It is calculated as the average "value" score of all remediation recommendations generated by the model. A higher score indicates more effective remediation recommendations.
        \[
        S_{R} = \frac{\sum_{s \in \mathcal{R}} \text{Value}(r) }{|\mathcal{R}|}
        \]
        where $\mathcal{R} = \mathcal{REM}(\mathcal{V}) = \mathcal{REM}(\mathcal{PEN}(\text{system}))$ represents the set of remediation strategies recommended by the Remediation Module ($\mathcal{REM}$).
        Using the average "value" score eliminates the influence of the number of recommendations on the score.
  \item \textbf{Remediation Cost Score ($C$)}: This score quantifies the resource intensity required to implement the remediation strategies, calculated as the average "cost" score of all remediation recommendations. A lower score indicates more efficient remediation.
        \[
        C = \frac{\sum_{s \in \mathcal{R}} \text{Cost}(r) }{|\mathcal{R}|}
        \]
        Using the average of the "cost" score also helps to mitigate the influence of the number of recommendations on the evaluation.
\end{itemize}

All of the three scores, by definition of the CVSS score and our definition of the ``cost'' and ``value'' score, are unitless.
Also, they fall in the range of $[0, 10]$, so that they can be directly compared.
The overall score for our model's performance is calculated using:
\[
S_{overall} = \frac{S_{D} + S_{R} - C}{3}
\]

This scoring system balances the importance of detection coverage rate, remediation effectiveness, and remediation efficiency, providing a holistic view of the model's capabilities.
By applying these metrics in the realistic settings of the Metasploitable2 environment, 
we aim to demonstrate the model's practicality in a scenario with diverse vulnerabilities.

\subsection{Numerical Results and Observations}
We assess the efficacy of our model, \textit{PenHeal}, 
by benchmarking it against two stable baseline models: PentestGPT and GPT-4. 
Our experimental setup involves conducting penetration testing and 
vulnerability remediation on the Metasploitable2 virtual machine using all three models. 
To mitigate the variability of LLM outputs, we repeat each experiment three times and average the scores. 
The penetration testing phase concludes once no additional vulnerabilities are detected by the tested model, ensuring a thorough evaluation of each model's detection capabilities.
For the experiments, we set a default cost budget of 4 units per vulnerability for \textit{PenHeal}, 
which balances the need for efficient remediation strategies with allowances for potentially resource-intensive solutions. 
To ensure fairness, the scores of the recommendations made by PentestGPT and GPT-4 are fed into the Evaluator to calculate their ``cost'' and ``value'' scores.
 
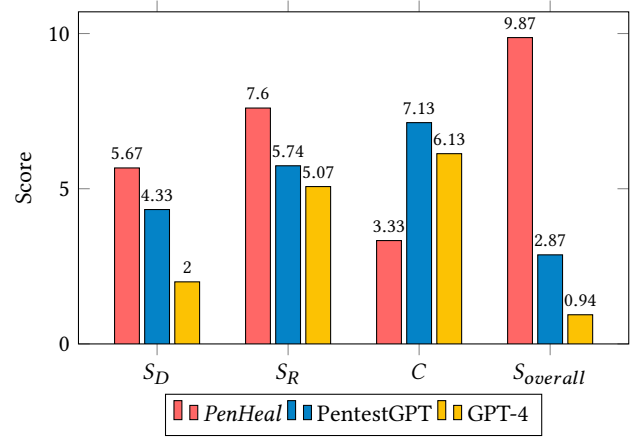
\begin{figure}[h]
  \centering
  \caption{Comparison of Scores Achieved by Different Models}
  \label{fig:score_comparison}
\begin{tikzpicture}
    \begin{axis}[
      ybar,
      symbolic x coords={$S_{D}$, $S_{R}$, $C$, $S_{overall}$},
      xtick=data,
      nodes near coords,
      nodes near coords align={vertical},
      ymin=0, ymax=10.7,
      ylabel={Score},
      legend style={at={(0.5,-0.15)},
      nodes near coords style={font=\footnotesize}, 
      anchor=north,legend columns=-1},
      width=0.5\textwidth,
      height=6cm,
      bar width=0.33cm,
      enlarge x limits=0.2,
      ]
  \addplot[fill=myred] coordinates {($S_{D}$, 5.67) ($S_{R}$, 7.60) ($C$, 3.33) ($S_{overall}$, 9.87)};
  \addplot[fill=myblue]  coordinates {($S_{D}$, 4.33) ($S_{R}$, 5.74) ($C$, 7.13) ($S_{overall}$, 2.87)};
  \addplot[fill=myorange]  coordinates {($S_{D}$, 2.00) ($S_{R}$, 5.07) ($C$, 6.13) ($S_{overall}$, 0.94)};
  
  \legend{\textit{PenHeal}, PentestGPT, GPT-4}
  \end{axis}
\end{tikzpicture}
\end{figure}

Figure~\ref{fig:score_comparison} shows the average score achieved by each model average over the 3 runs. 
The results show that \textit{PenHeal} improves the detection of vulnerabilities by 31\% compared to baseline models, as evidenced by a Detection Coverage Score ($S_{D}$) of 5.67. Besides, \textit{PenHeal} boosts the effectiveness of remediation strategies by 32\% compared to PentestGPT with a remediation effectiveness score ($S_{R}$) of 7.6. In terms of cost score ($C$), \textit{PenHeal} also reduces the associated costs by 46\% compared to baseline models. The overall performance score ($S_{overall}$) of 9.87 demonstrates that PenHeal is highly efficient and effective across all measured parameters.

It is worth noting that when conducting the experiments with PentestGPT and GPT-4, 
human operator is required to participate in order to extract information from the verbose outputs, make decisions, and facilitate the progression of tasks. 
Sometimes, the two models may stop after successfully exploiting only a vulnerability or working on a failed attempt non-stop, and the human operator has to step in
and prompt them to move on to the next task.
However, for \textit{PenHeal}, the human operator only needs to provide the IP address of the victim machine, and the rest of the process is fully automated.

\subsection{Ablation Study}

\begin{figure}[h]
  \centering
  \caption{Ablation Study of \textit{PenHeal}}
  \label{fig:ablation_study}
  \begin{tikzpicture}
      \begin{axis}[
          ybar,
          legend style={at={(0.5,-0.20)},
              anchor=north,legend columns=2},
          ylabel={Score},
          symbolic x coords={$S_{D}$, $S_{R}$, $C$, $S_{overall}$},
          xtick=data,
          bar width=0.3cm,
          nodes near coords,
          nodes near coords align={vertical},
          nodes near coords style={font=\footnotesize},
          ymin=0, ymax=10.7,
          x tick label style={rotate=45, anchor=east},
          width=0.5\textwidth,
          height=7cm,
          enlarge x limits=0.17,
      ]
      \addplot[fill=myred] coordinates {($S_{D}$, 5.67) ($S_{R}$, 7.60) ($C$, 3.33) ($S_{overall}$, 9.87)};
      \addplot[fill=myblue] coordinates {($S_{D}$, 1.67) ($S_{R}$, 8.13) ($C$, 3.00) ($S_{overall}$, 6.80)};
      \addplot[fill=myorange] coordinates {($S_{D}$, 2.33) ($S_{R}$, 8.2) ($C$, 2) ($S_{overall}$, 8.53)};
      \addplot[fill=mycolor2] coordinates {($S_{D}$, 5.67) ($S_{R}$, 6.21) ($C$, 6.40) ($S_{overall}$, 5.48)};
      
      \legend{\textit{PenHeal}, \textit{PenHeal}-noCounterfactual, \textit{PenHeal}-noInstructor, \textit{PenHeal}-noEvaluator}
      \end{axis}
\end{tikzpicture}
\end{figure}
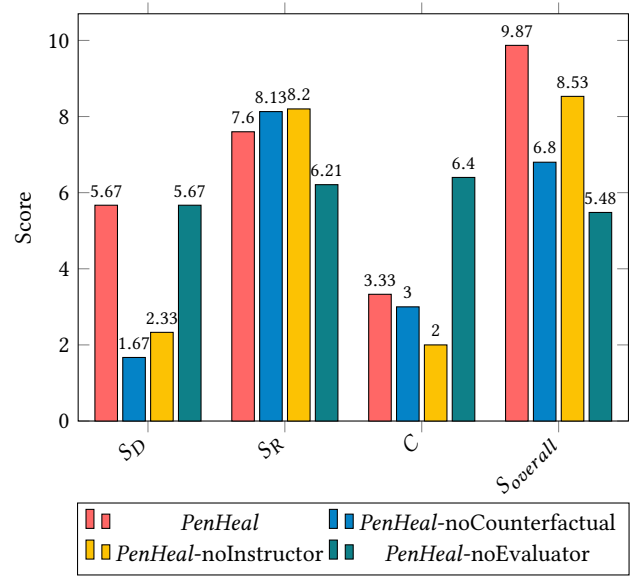

To understand the contribution of some components in our model, we conduct an ablation study by respectively removing 
the following in our model: the Counterfactual Prompting in the Planner, the Instructor, the Evaluator. 
As a result, we have the following variants of our model: \textit{PenHeal}-noCounterfactual, \textit{PenHeal}-noInstructor, \textit{PenHeal}-noEvaluator.
We compare these three variants with the original model and evaluate their performance using the same benchmark as before.
The results are shown in Figure~\ref{fig:ablation_study}.

The ablation study reveals the critical role of Counterfactual Prompting in the Planner, which significantly boosts the Detection Coverage Score ($S_{D}$) of \textit{PenHeal} to 5.67. Without this feature, as observed in the \textit{PenHeal}-noCounterfactual variant, the model tends to repeat the same attack paths even after tasks are marked as completed, leading to a drop by 71\% in $S_{D}$. 


\begin{tcolorbox}[colback=yellow!10!white,colframe=yellow!50!black]
\textbf{Key Message \#1}: LLMs require an internal ``stimulus'' (facilitated by the Counterfactual Prompting) to explore various possibilities, thereby enabling the discovery of multiple vulnerabilities.
\end{tcolorbox}

The Instructor's contribution to detection efficiency is also notable. It equips the Executor with essential information to generate accurate commands, enhancing the Detection Coverage Score ($S_{D}$). The absence of the Instructor, as seen in the \textit{PenHeal}-noInstructor variant, often results in incorrect command generation and hence a significant drop in $S_{D}$ by 59\%. 

\begin{tcolorbox}[colback=yellow!10!white,colframe=yellow!50!black]
\textbf{Key Message \#2}: LLMs depend on external knowledge (supported by the Instructor component) to generate precise commands and effectively identify vulnerabilities.
\end{tcolorbox}

Lastly, the Evaluator significantly impacts the quality of remediation recommendations. Removing the Evaluator leads to the \textit{PenHeal}-noEvaluator variant producing many costly or ineffective recommendations, such as unnecessary service shutdowns or security audits, reflected in the highest Cost Score ($C$) of 6.40 and the lowest overall score of 5.48.

\begin{tcolorbox}[colback=yellow!10!white,colframe=yellow!50!black]
\textbf{Key Message \#3}: LLMs need a ``filter'' (supported by the Evaluator) to generate remediation suggestions that are effective, efficient, and accommodated to user preference.
\end{tcolorbox}

\subsection{Demonstration of Model Outputs}
In this subsection, we will showcase some textual outputs of the model in the runtime for a more intuitive understanding.

\subsubsection{Demonstration of the Attack Plan}
\label{sec:attack_plan_demo}
\begin{figure}[ht]
  \centering
  \includegraphics[width=\columnwidth]{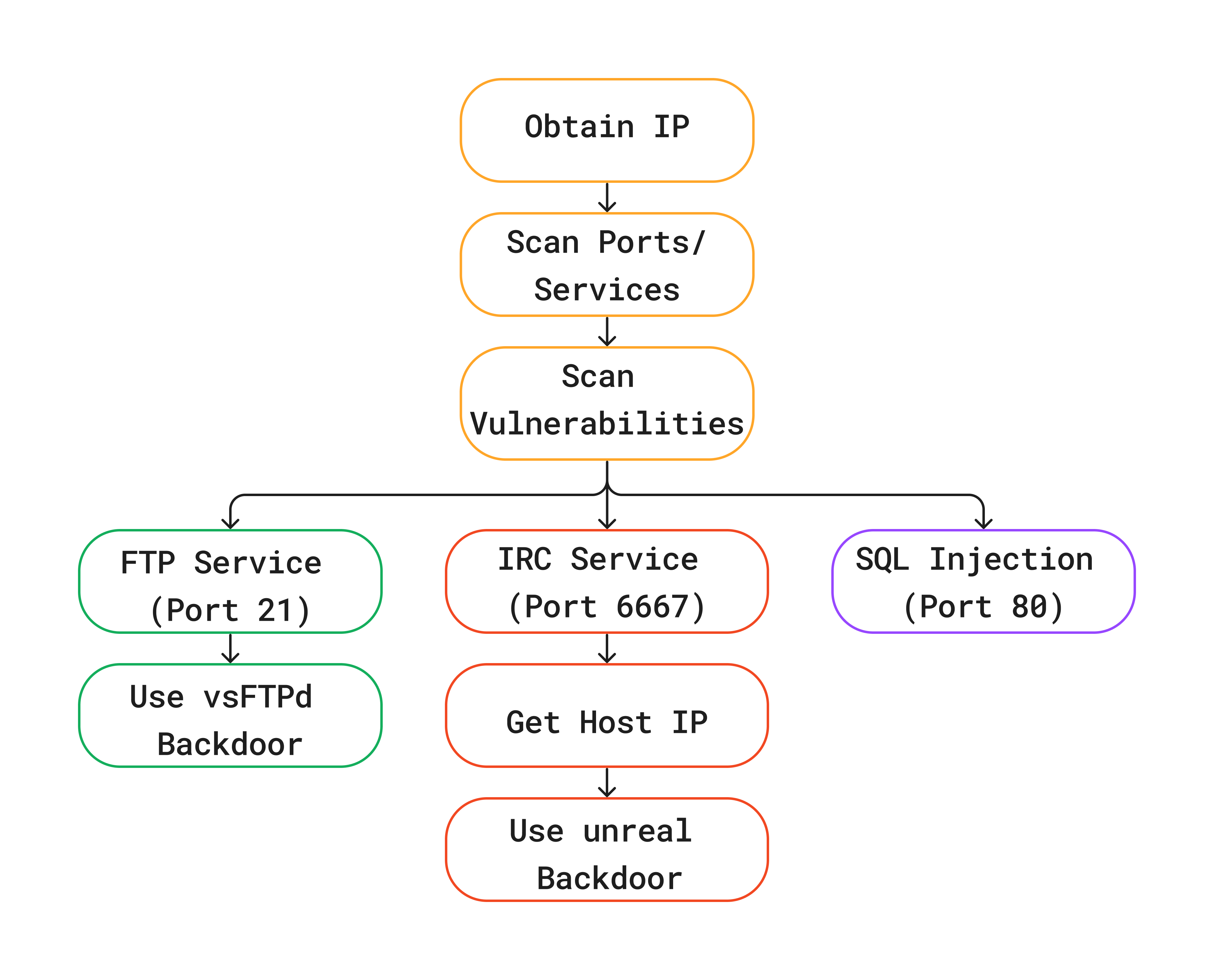}
  \caption{A visualization of the attack plan in tree structure, 
          with each node representing a specific task. The orange block indicates Reconnaissance tasks, while the green, red, and purple blocks represent exploits on Ports 21, 6667, and 80, respectively.}
  \label{fig:attack_plan_tree}
\end{figure}

An excerpt of the attack plan generated in the run time is shown in Figue~\ref{fig:attack_plan_sequence}, and its visualization is shown in Figue~\ref{fig:attack_plan_tree}.
We can see that the Planner is adept at searching for useful information in the Reconnaissance phase to inform its attack strategy. Additionally, when an attack (such as the exploit on the vsFTPd backdoor) succeeds, the Planner efficiently marks the task as completed and promptly initiates a new task. Furthermore, if an attack fails, the Planner can determine the cause of the failure and intelligently add subtasks to gather additional information for a retry, demonstrating its adaptability.

\subsubsection{Demonstration of the Instructor's Excerpt}
\label{sec:instructor_demo}
The Instructor is responsible for providing guidance to the Executor by retrieving relevant information from the external knowledge base.
An excerpt in response to the task ``\texttt{Exploit Samba smbd 3.X - 4.X on ports 139 and 445}'' is demonstrated in Listing~\ref{lst:retrieved_excerpt}, where the Instructor effectively retrieves and supplies crucial information to the Executor. 

\begin{lstlisting}[caption=Example of Retrieved Excerpt, label=lst:retrieved_excerpt]
  Excerpt 1: Employ the use command after the 
  exploit name: 
  msf > use exploit/multi/samba/usermap_script

  This module exploits a command execution 
  vulnerability in Samba versions 3.0.20 through 
  ...
\end{lstlisting}

\subsubsection{Demonstration of the Remediation recommendations}
\label{sec:recommendation_demo}
The recommendations given by the Adviser are multifaceted, and it may not be easy for non-expert users to decide which one to follow, 
as can be seen in Listing~\ref{lst:Advisor_recommendations}. 
This may not be practical in real life, as the users may not want to shut down the service or stop the system, 
because it may disable the service that is essential for the business.

\begin{lstlisting}[caption=Output of the Advisor on Samba Vulnerability, label=lst:Advisor_recommendations]
  1. Update Samba using ...
  2. Perform regular security audits ...
  3. Shut down the Samba service ...
  4. Configure firewall using ...
  ...
\end{lstlisting}

However, after being processed by the Evaluator and the Group Knapsack Algorithm, the remediation recommendations become more targeted and effective. For instance, when the only detected issue is a Samba vulnerability, the budget is set to 3 and the default cost evaluation metric, the recommendations from the Advisor are narrowed down as illustrated in Listing~\ref{lst:Advisor_recommendations_narrowed}. 
In this case, the recommendation of either updating Samba or configuring the firewall is adopted randomly, 
as they can both directly address the vulnerability with little effort.
Conversely, the recommendation for regular security audits is rejected because they do not directly tackle the issue, and shutting down the Samba service is avoided due to its high disruptiveness and impracticality in operational environments.

\begin{lstlisting}[caption=Output of the Advisor on Samba Vulnerability after Evaluation, label=lst:Advisor_recommendations_narrowed]
  1. [Adopted] Update Samba using...
    Cost: 2.0, Value: 9.0
  2. [Discarded] Perform regular security audits...
    Cost: 2.0, Value: 3.0
  3. [Discarded] Shut down the Samba service...
    Cost: 10.0, Value: 9.0
  4. [Adopted] Configure firewall using...
    Cost: 2.0, Value: 9.0
  ...
\end{lstlisting}

\section{Discussion}

\subsection{Answer to the Research Questions}
Our experimental results validate that our model, \textit{PenHeal}, can effectively automate penetration testing that discovers multiple vulnerabilities without human involvement. As detailed in Figure~\ref{fig:score_comparison}, \textit{PenHeal} surpasses baseline models such as PentestGPT and GPT-4 in detection coverage ($S_{D}$), identifying approximately 60\% of the vulnerabilities in the Metasploitable2 environment. Therefore, in response to Research Question 1 (RQ1), we can conclusively state that LLMs can indeed automate the discovery of various vulnerabilities in a target system, performing effectively without the need for human intervention.

\begin{tcolorbox}[colback=yellow!10!white,colframe=yellow!50!black]
\textbf{Key Message \#4}: LLMs can be used to automate the discovery of various vulnerabilities in a target system, performing effectively without the need for human intervention.
\end{tcolorbox}
The performance of our Remediation Module in automating vulnerability remediation also offers a positive answer to RQ2. 
The effectiveness and efficiency of the remediation recommendations generated by \textit{PenHeal} (quantified by $S_{R}$ and $C$)
turned out to be significantly higher than those produced by PentestGPT and GPT-4.
This demonstrates an enhanced capability over existing models, 
highlighting LLMs's potential in vulnerability remediation.

\begin{tcolorbox}[colback=yellow!10!white,colframe=yellow!50!black]
\textbf{Key Message \#5}: LLMs can be used to generate remediation strategies that are effective, efficient and tailored to user preference.
\end{tcolorbox}

\subsection{Limitations and Future Work}

While \textit{PenHeal} has shown promising results in automating penetration testing and vulnerability remediation, it primarily serves as a proof of concept to illustrate the potential applications of LLMs in cybersecurity automation. Therefore, our model needs next-step refinement before becoming a fully mature, plug-and-play solution.

To start with, similar to AUTOATTACKER \cite{xu2024autoattacker}, our model sometimes hallucinates, leading to the generation of incorrect commands, such as using non-existent modules in Metasploit or failing to adhere to task requirements. We envision this issue can be mitigated with the advancement in LLM technique. 

The effectiveness of our model is also highly dependent on the specific attack tools included in its knowledge base. This dependency limits its ability 
to conduct penetration tests on systems that are more sophisticated or have restricted access to these tools. Therefore, future work could focus on enhancing the model's integration of diverse cybersecurity tools to widen its applicability across different network settings. 


Moreover, while the model performs well on a single machine, its efficacy on multiple machines within a network may vary. Due to time constraints, our testing has been limited to the Metasploitable2 machine, without the opportunity for broader experimentation. Future work should include extensive testing across various network scenarios to enhance the model's applicability.

Finally, the model performs well in formulating remediation strategies, but it currently does not execute these strategies directly, because it operates from the attacker's host machine. Therefore, efforts could be made to enable the model to not only suggest but also implement remediation actions, which could involve creating a secure bridge between the attacker's and victim's systems. By addressing this limitation, the model could provide end-to-end automation of penetration testing and vulnerability remediation, achieving 
a self-healing effect.

\subsection{Ethical Considerations}

The deployment of a model able to conduct automated attacks may raise security and privacy concerns. 
Indeed, our model is intentionally prompted to avoid any form of exploitation following discovery, 
which can reduce the risk of accidental damage. 
However, if not properly secured, the model itself could be exploited by malicious hackers to discover and exploit vulnerabilities rather than remediate them. Therefore, cautious monitoring and usage of the model are necessary.

\section{Conclusion}

In this work, we introduce a novel model, \textit{PenHeal}. 
It uses LLMs to automate the entire penetration testing process that could discover multiple vulnerabilities, 
and generate high-quality remediation recommendations with minimal human intervention. 
The pipeline starts with the user providing the IP address of the victim machine and ends with the model generating a selected list of remediation recommendations according to the user's preference and budget. 
Experimental results show that \textit{PenHeal} outperforms baseline models, PentestGPT and GPT4, in terms of detection coverage, remediation effectiveness, and cost efficiency,
demonstrating the feasibility of deploying LLMs in advanced cybersecurity tasks.

Although our model needs further updates to become a fully mature, plug-and-play solution, it contributes to filling the
research gap of leveraging LLMs for penetration testing for multiple vulnerabilities and their remediation with full automation.
We hope it could inspire future research on many aspects, for instance, 
getting rid of the model's dependency on specific attack tools, enhancing its knowledge of vulnerability remediation, 
and achieving end-to-end automation including conducting vulnerability remediation.

\printbibliography
\appendix
\section{System Prompts}
\subsection{Planner System Prompt}
\label{sec:planner_system_prompt}
\begin{lstlisting}[caption=Planner System Prompt, label=lst:Planner_system_prompt]
  As the chief director of penetration testing, 
  you are responsible for high-level planning. 
  Please follow these guidelines:
  (1) Task Structure: Organize tasks in a layered 
      sequence (e.g., 1, 1.1, 1.1.1). 
  (2) Assign a status to each task: to-do, 
      completed, or failed. Update this status and 
      include succinct results based on 
      the latest outcomes.
  (3) If uncertain about a task or the previous 
      outcomes indicate that more info is 
      required, you are encouraged to add subtasks 
      to acquire more information
      (e.g., search in Metasploit).
  (4) Choose the next task based on the latest 
      outcomes and the current attack plan. You 
      can either start a new task (marked as to-do) 
      or continue with the current one.
  ......
\end{lstlisting}

\subsection{Executor System Prompt}
\label{sec:executor_system_prompt}
\begin{lstlisting}[caption=Executor System Prompt, label=lst:Executor_system_prompt]
  ... you are responsible for generating commands 
  given the task assigned by your boss...
  (1) Give your command(s) in a format starting 
      and ending with "$" for easy parsing.
  (2) If the command is too long or complex, 
      break it down into multiple commands.
  (3) Only give a sequence of commands if they 
      are logically connected and follow a 
      sequence that makes sense.
  (4) If you want to use Metasploit, please write 
      in this format: "msfconsole: [command]".
  (5) Assume that you are operating on a Kali 
      Linux machine. All tools are pre-installed. 
      You are given the permission to use all of 
      them.
  ......
\end{lstlisting}



\subsection{Estimator System Prompt}
\label{sec:estimator_system_prompt}
\begin{lstlisting}[caption=Estimator System Prompt, label=lst:Estimator_prompt_template]
  ... you are responsible for producing a vector 
  string for CVSS calculator...
  (1) A vector string is of the form 'CVSS:3.0/AV:
  [value]/AC:[value]/PR:[value]/UI:[value]
  /S:[value]/C:[value]/I:[value]/A:[value]'.
  (2) Attack Vector(AV): The way the vulnerability 
      is exploited. Choose from: Network (N),
      Adjacent (A), Local (L), Physical (P).
  (3) Attack Complexity(AC): The conditions beyond
      the attacker's control that must exist to 
  ......
\end{lstlisting}

\subsection{Advisor System Prompt}
\label{sec:advisor_system_prompt}
\begin{lstlisting}[caption=Advisor System Prompt, label=lst:Advisor_prompt_template]
  ... you are responsible for generating remediation 
  strategies for the vulnerabilities identified.
  (1) Ensure that the remediation recommendations 
      are actionable and effective in mitigating 
      the identified risks.
  (2) Specific commands are preferred over general 
      recommendations. For example, "sudo apt 
      update" is better than "Update the system".
  (3) Adding protection is preferred over shutting 
      down services. For example, suggesting 
      enabling firewall and giving commands 
      better than "Shut down the service".
  ......
\end{lstlisting}

\subsection{Evaluator System Prompt}
\label{sec:evaluator_system_prompt}
\begin{lstlisting}[caption=Evaluator System Prompt, label=lst:Evaluator_prompt_template]
  You have already discovered these vulnerabilities: 
  {Vulnerabilities}. If the following remediation 
  approach is adopted, to what extent can it 
  mitigate the risks? Give the score based on the 
  rule: {Definition of "value"}. Give the score 
  based on their cost: {Definition of "cost"}.
\end{lstlisting}

\section{Attack Plan Script}
\label{sec:attack_plan_script}
\begin{figure}[ht]
  \centering
  \includegraphics[width=\columnwidth]{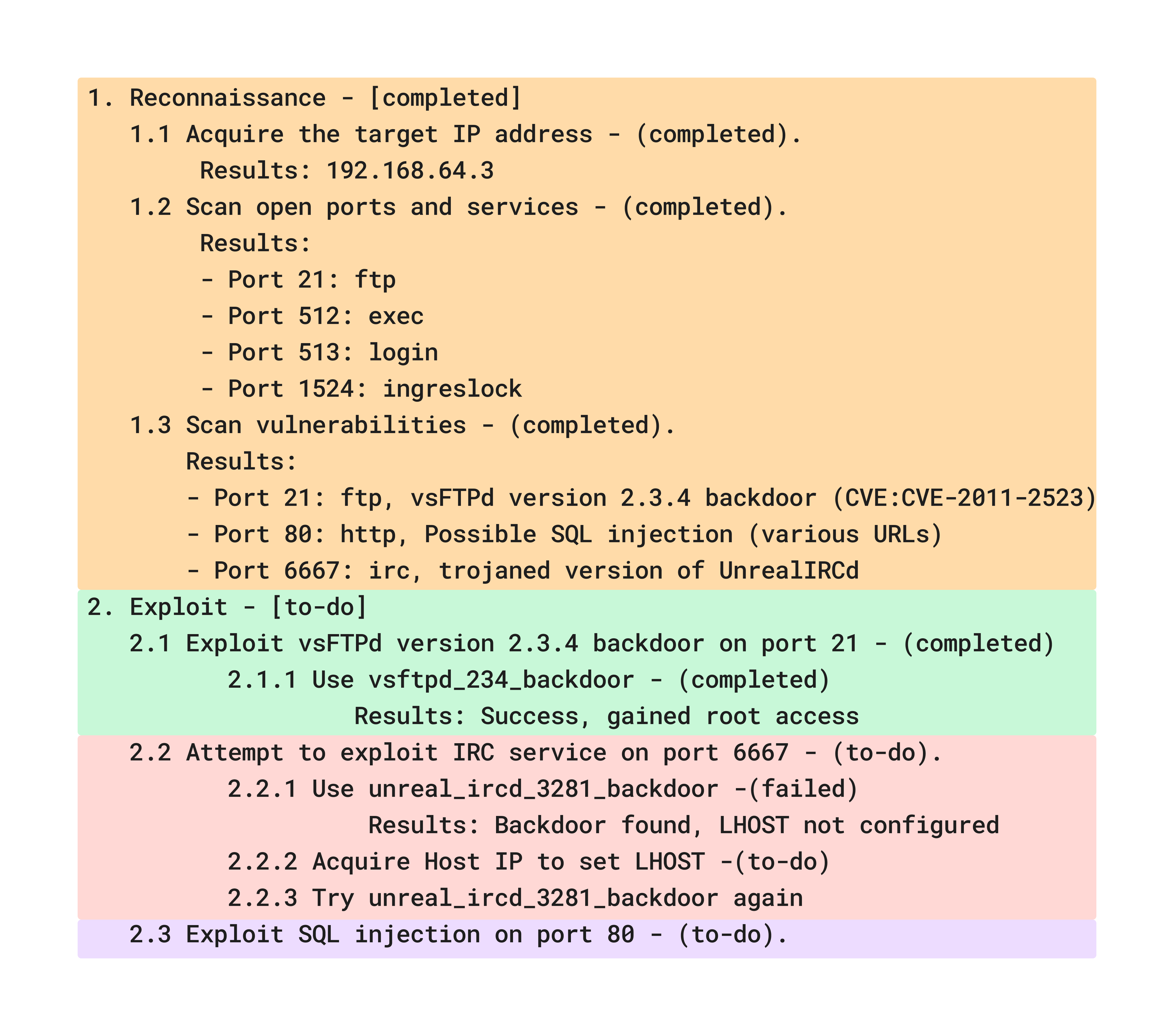}
  \caption{A sample attack plan in a layered structure.}
  \label{fig:attack_plan_sequence}
\end{figure}

\section{Group Knapsack Algorithm}
\label{sec:group_knapsack}
The Group Knapsack Algorithm is a variant of the classic Knapsack Problem. 
The goal is the same: to maximize the total value of items selected while keeping the total weight below a certain threshold.
However, in the Group Knapsack Algorithm, the items are divided into groups, and only one item from each group can be selected. 
\begin{lstlisting}[caption={Pseudo-code of the Group Knapsack problem}, label=lst:groupknapsack]
  function GroupKnapsack(groups, C):
    // groups: an array of various groups of items
    // C: the knapsack capacity
    dp = array of size (C+1) with zeros

    for each group in groups:
      for j from C down to 0:
        for each item in group:
          if j >= item.weight:
            dp[j]=max(dp[j],
            dp[j - item.weight] + item.value)
    return dp[C]
\end{lstlisting}

\end{document}